\documentclass[12pt,nohyper,notoc]{JHEP}
\input epsf
\epsfverbosetrue
\usepackage{amsmath,euscript,array,amssymb,cite} 
\newcommand{\startappendix}{
\setcounter{section}{0}
\renewcommand{\thesection}{\Alph{section}}}

\newcommand{\Appendix}[1]{
\refstepcounter{section}
\begin{flushleft}
{\large\bf Appendix \thesection: #1}
\end{flushleft}}

\def\N{{\cal N}}

\def\Tr{{\rm Tr}}
\def\sst{\scriptscriptstyle}

\def\Dbarslash{\,\,{\raise.15ex\hbox{/}\mkern-12mu {\bar\D}}}
\def\Dslash{\,\,{\raise.15ex\hbox{/}\mkern-12mu \D}}
\def\delslash{\,\,{\raise.15ex\hbox{/}\mkern-9mu \partial}}
\def\delbarslash{\,\,{\raise.15ex\hbox{/}\mkern-9mu {\bar\partial}}}

\def\Z{{\EuScript Z}}

\def\hf{{\textstyle{1\over2}}}

\newcommand{\EQ}[1]{\begin{equation} #1 \end{equation}}

\newcommand{\SP}[1]{\begin{equation}\begin{split} #1 \end{split}\end{equation}}

\def\beqa{\begin{eqnarray}} 
\def\eeqa{\end{eqnarray}} 
\def\beq{\begin{equation}} 
\def\eeq{\end{equation}} 
\def\R{\mbox{\rm I\kern-.18em R}} 
\def\Rq{\R^4} 
\def\P{\mbox{\rm I\kern-.18em P}} 
\def\uno{\mbox{1 \kern-.59em {\rm l}}} 
\def\Z{{Z \kern-.45em Z}} 
\def\Q{{\kern .1em {\raise .47ex \hbox{$\scriptscriptstyle |$}} 
\kern -.35em {\rm Q}}} 
 
\def\Tr{\mbox{\rm Tr}}

\def\p{\partial}

\def\ie{{\it i.e. }} 
 
\font\mybb=msbm10 at 12pt

\def\bb#1{\hbox{\mybb#1}} 
\def\Z {\bb{Z}}

\title{Wilsonian Effective Actions and the IR/UV Mixing
in Noncommutative Gauge Theories}

\author{Valentin V.~Khoze and Gabriele Travaglini\\
Department of Physics and IPPP, University of Durham,
Durham, DH1 3LE, UK\\
E-mail: 
{\tt valya.khoze@durham.ac.uk}, {\tt gabriele.travaglini@durham.ac.uk}}

\abstract{Using background field perturbation theory we study Wilsonian
effective actions of noncommutative gauge theories with an arbitrary matter
content. We determine the Wilsonian coupling constant and the gauge boson
polarization tensor as functions of the momentum scale $k$ at the one-loop level
and study their short-distance behaviour as $\theta \cdot k\to 0$, where
$\theta$ is the noncommutativity parameter. The mixing between the
short-distance and the long-distance degrees of freedom characteristic of
noncommutative field theories violates the universality of the Wilsonian action and
leads to IR-singularities. We find, in agreement with known results,
that the quadratic IR divergences cancel in supersymmetric gauge theories.
The logarithmic divergences disappear in mass-deformed ${\cal N}=4$ theories,
but not in other finite ${\cal N}=2$ theories. 
\newline
We next concentrate on finite ${\cal N}=2$ and mass-deformed ${\cal N}=4$
supersymmetric $U(1)$ gauge theories with massive hypermultiplets. The Wilsonian
running coupling exhibits a non-trivial threshold behaviour at and well below
the noncommutativity scale $1/\sqrt{\theta}$, eventually becoming flat 
in the extreme infrared in ${\cal N}=4$ theories, but not in ${\cal N}=2$ theories.
This is interpreted as the (non)-existence of a non-singular commutative limit
where the theory is described by a commutative ${\cal N}=2$ pure $U(1)$ theory.
We expect that our analysis of finite theories is exact to all orders in
perturbation theory.} 

\preprint{{\tt hep-th/0011218}}

\begin{document}

\section{Introduction}\label{sec:S1}

Gauge theories on noncommutative spaces have recently attracted
much attention for their
applications to string and matrix theories \cite{CDS,DHull,SWnc}.
Supersymmetric gauge theories arise as the
low-energy description of open strings ending on D-branes
in the presence of a constant $B$-field
which gives rise to space-noncommutativity,
\EQ{[x^\mu, x^\nu]=i\theta^{\mu\nu} \ . \label{nctheta}}
In this paper we analyze the field theory
dynamics of noncommutative gauge theories 
in four spacetime dimensions and in the
presence of a generic matter field content. 
For related recent work see 
\cite{Matusis,Minwalla,Filk,Sheikh-Jabbari,Martin,
Krajewski,Arefeva,Hayakawa,Hayakawa2,Ferrara,Terashima:2000xq,
Gracia-Bondia,Girotti,Bichl,
Armoni,Ambjorn,Semenoff} and references therein.
Noncommutative field theories can be defined by replacing the ordinary
products of fields in the Lagrangians of their commutative counterparts 
by the star-products
\EQ{(\phi \star \chi) (x) \equiv \phi(x) e^{{i\over 2}\theta^{\mu\nu}
\stackrel{\leftarrow}{\partial_\mu}
\stackrel{\rightarrow}{\partial_\nu}}  \chi(x) \ . \label{stardef}}
In this way noncommutative theories can be viewed as field theories on 
 ordinary commutative spacetime. For example,  the noncommutative pure
gauge theory action is
\EQ{
S_{\rm YM} [A] = -{1\over 2g^2}\int d^{4} x \ \Tr ( F_{\mu \nu}\star  F_{\mu \nu}
 ) \ , \label{pureym} 
}  
where
the field strength is defined
as $F_{\mu \nu} = \p_\mu A_\nu - \p_\nu A_\mu +  
[A_\mu , A_\nu]_{\star}$.
If we Taylor-expand 
the star-products in \eqref{pureym} we obtain the action
of the standard commutative theory plus
an infinite number of higher-derivative
terms. At an energy-scale below the noncommutativity scale, $k^2\ll 1/\theta$,
the higher-derivative terms correspond to irrelevant operators.
One would normally expect that in the infrared one can simply drop
all the effects due to irrelevant operators. 
This would imply that the noncommutative and the corresponding commutative
theories belong to the same universality class, \ie\ in the infrared their
behaviour is identical. Classically the two theories are, in fact, identical
in this regime. But it turns out \cite{Minwalla,Matusis} that
this naive universality is invalidated at the quantum level due to a curious
mixing between the short-distance and the
long-distance modes in the loop expansion of noncommutative
theories. 

Standard commutative theories are known to be universal, \ie\ their
UV-modes are decoupled from the IR-modes. This universality apparently
goes wrong in noncommutative theories. The reason for this lies
in the UV-properties of these theories.
It is believed that the
noncommutative theories are UV-renormalizable (when their commutative
counterparts are) contrary to the naive expectations about
irrelevant higher-derivative operators.
In perturbation theory \cite{Filk, Minwalla}
the loop integrals in the planar diagrams
of the noncommutative theory are exactly the same as in the commutative 
counterpart. The non-planar diagrams, however, are multiplied by phase
factors of the form $e^{ik\cdot\theta\cdot p}$, 
where $k$ are external momenta,
and $p$ are loop momenta. These oscillating phases improve the 
UV-convergence of the non-planar diagrams and typically render them finite.
This is basically the reason for the UV-renormalizability of the 
noncommutative theory\footnote{Note, that if we had Taylor-expanded the
star-products, we would have ended up with the Taylor expansion for the phase
factors rendering each term in the expansion progressively more singular in the
UV region of loop momenta.}. But in the IR limit of the external
momenta, $k\to 0$, the phases vanish and the non-planar
diagram diverge, but this divergence is now interpreted 
as an IR-divergence\cite{Minwalla, Matusis}.
This is the origin of the IR/UV mixing in the noncommutative theories
and it leads to the breakdown of universality. 

In this paper we study the IR/UV mixing, (non)-universality 
and the IR commutative
limits $\theta k \to 0$ of generic supersymmetric and non-supersymmetric
gauge theories in the Wilsonian approach, which is well suited
for addressing the issues of the IR dynamics. We evaluate and discuss 
the Wilsonian running coupling and the gauge boson polarization
tensor as functions of the Wilsonian momentum scale $|k|$ at the one-loop level.
The IR/UV mixing violates universality of the Wilsonian action
and leads to the IR-singularities: $(\theta\cdot k)^{-2}$ divergences 
in the polarization tensor and new $\log(k\cdot\theta\cdot k)$
divergences in the coupling
in addition to the standard $\theta$-independent running.
We find, in agreement with known results \cite{Matusis}, that
the $(\theta \cdot k)^{-2}$ divergences cancel
in all supersymmetric theories. The $\log(k\cdot\theta\cdot k)$ 
divergences disappear in finite $\N=4$ supersymmetric
theories (with or without supersymmetry breaking mass terms)
in agreement with \cite{Matusis}, but not in finite $\N=2$ theories
with fundamental hypermultiplets. 

We analyze in detail the running of the Wilsonian coupling in supersymmetric
noncommutative $U(1)$ theories with asymptotically free and vanishing
microscopic $\beta$ functions. All these theories
are not universal, and in the commutative limit their dynamics
are different from the naive expectations. 
Equations (\ref{45}) and (\ref{25}) summarize 
our results for  the running of the Wilsonian coupling constants 
in the noncommutative {\it finite} supersymmetric theories, 
as shown  in Figures 1 and 2 respectively. 
Let us first describe the results of
our analysis of the $\N=4$ theory softly broken
to $\N=2$ by the adjoint hypermultiplet mass term. From now on this
will be referred to as the $\N=2^*$ theory. Figure 1 shows the running of the
Wilsonian coupling of this theory in four different regions.
The first one is the UV region, $k>>m$, where $m$ is the mass of the adjoint 
chiral multiplets. Here the coupling is flat as it should be
for the $\N=4$ noncommutative theory. In the second region,
$\theta^{-1/2}<< k<< m$, the coupling is asymptotically free, again in agreement
with expectations, since in this regime the massive hypermultiplets are
integrated out and the theory is simply a pure noncommutative $\N=2$ $U(1)$
 (which is asymptotically free). In the third region, $k<\theta^{-1/2}$,
the coupling exhibits a non-trivial threshold behaviour at and well below the noncommutativity
scale $\theta^{-1/2}$, eventually becoming flat in the extreme infrared,
$k<<\theta^{-1/2}$ (in the region four). 
This is interpreted as the existence of a non-singular commutative limit
where the theory is described by a commutative ${\cal N}=2$ pure $U(1)$ theory
which indeed has $\beta=0$.

\vfil\eject

\epsfxsize=18cm 
\centerline{\epsffile{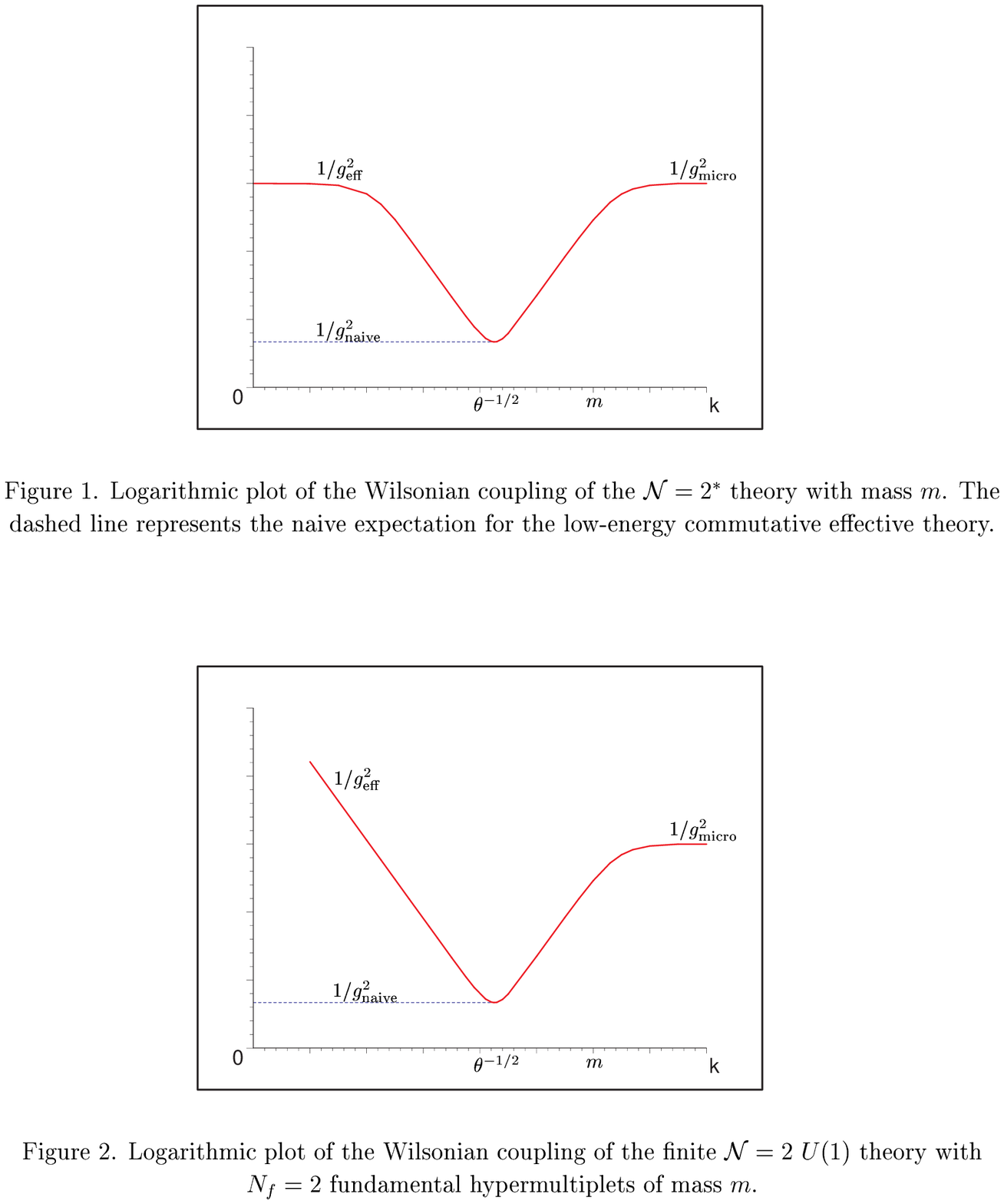}}
\vfil\eject

The Wilsonian coupling of the noncommutative finite 
$\N=2$ $U(1)$ theory with $N_{\bf f}=2$ fundamental hypermultiplets with mass $m$
as a function of the momentum scale is shown in Figure 2.
Here the Wilsonian coupling exhibits the screening behaviour 
all the way below the noncommutativity
scale $1/\sqrt{\theta}$. This has no resemblance to a commutative 
${\cal N}=2$ pure $U(1)$ theory, which is supposed to be flat.
Somehow, in the IR the theory becomes commutative, but not the same flat
commutative one as one would expect in the classical limit.
We interpret this fact as a loss of universality in the IR regime.

Figures 1 and 2 show that, remarkably, the effective coupling 
$g_{\rm eff}$ never
explodes ($1\over g_{\rm eff}^2$ does not reach zero)
for not too large values of the microscopic coupling 
$g_{\rm micro}$. 

The organization of the rest of the paper is as follows.
In Section 2 we introduce the background field method for noncommutative
gauge theories which is used for calculating the Wilsonian effective
action in perturbation theory. In Section 3 we present the Feynman rules
which are used in Sections 4 and 5 to compute
the 1-loop determinants arising from integrating out the fluctuating 
fields. Then the Wilsonian running coupling constant 
and the gauge boson polarization
tensor are computed as functions of the momentum scale $k$.
We also discuss the simplifications which occur in
supersymmetric theories.
In Section 6 
we analyze the Wilsonian flow of finite massive supersymmetric gauge
theories down to the infrared.
To fully exploit our  formalism, 
we compute in Section 7 the evolution of the Wilsonian coupling constant 
for asymptotically free pure $\N=1$ and $\N=2$ Super Yang-Mills theories.
Finally, in the Appendix we present an alternative simple 
derivation of the microscopic
$\beta$ function for noncommutative $U(N)$ gauge theories 
counting zero modes of noncommutative instantons.
 
\bigskip

\centerline{\it Note on Conventions}

{\parindent=0 pt
Throughout the paper we work in Euclidean space. 
Our $\sigma_\mu$  and $\bar{\sigma}_\mu$ matrices are defined as
$\sigma_\mu=(i\sigma^m , \uno_{2\times 2})$ and  
$\bar{\sigma}_\mu=(-i\sigma^m , \uno_{2\times 2})$ in terms of the
three Pauli matrices $\sigma^m.$   
We will also use  
$\sigma_{\mu \nu}= {1\over 2} (\sigma_\mu \bar{\sigma}_\nu-\sigma_\nu \bar{\sigma}_\mu)=  
i \eta^{a}_{\mu \nu}\sigma^a$, and  
$\bar{\sigma}_{\mu \nu}=  
{1\over 2} (\bar{\sigma}_\mu \sigma_\nu-\bar{\sigma}_\nu \sigma_\mu)=  
i \bar{\eta}^{a}_{\mu \nu}\sigma^a$, where $\eta^{a}_{\mu \nu}$ and  
$\bar{\eta}^{a}_{ \mu \nu}$ 
are  the 't Hooft symbols \cite{'tHooft:1976}. 

For the general discussion of the gauge group $U(N)$ in Section 2
and in the Appendix
we use anti-hermitian gauge-group generators $t^A$ with the normalization 
$\Tr (t^A t^B ) = - {\delta^{AB} \over 2}$.  
Hence the generator of the $U(1)$ component of $U(N)$ is 
$t^0 = {1 \over i\sqrt{2N}}$. 

In Sections 3-7 we concentrate on the $U(1)$ gauge group
and change the
normalization of the $U(1)$ generator to $t^0 = {1 \over i}$.}

\vfil\eject

\section{The Background Field Method for Noncommutative Gauge Theories} 
In this Section we discuss the set-up of the method for  
noncommutative $U(N)$ gauge theories.

We start off decomposing the gauge field $A_\mu$ into a background field
$B_\mu$ and a fluctuating quantum field $N_\mu$, 
\EQ{
A_\mu= B_\mu + N_\mu 
\ \ . 
}
$N_\mu$ is treated as a highly virtual field 
with momenta above the Wilsonian scale while 
the background field is taken to be slowly varying, 
but still fully noncommutative. We are interested in the Wilsonian
effective action $S_{\rm Weff}(B)$ which is obtained by functionally
integrating over the fluctuating fields. For the pure gauge theory
we have (schematically)
\EQ{ \exp[-S_{\rm Weff}(B)] \ = \ \int {\cal D}N \ \exp[-S_{\rm YM} (A)] \ .
\label{Wdef}}

The noncommutative pure Yang-Mills action is given by \eqref{pureym}.
For a {\it fixed} background field the Yang-Mills action
 has a gauge symmetry
for the fluctuating field $N_\mu$.
Hence we need to fix the gauge and we will do this by adding
to \eqref{pureym}
the gauge-fixing functional \cite{Weinberg} 
\EQ{
S_{\rm g.f.} = -{1 \over g^2}\int d^{4} x \ 
\Tr \left( \bigl(D_{\mu}(B) N_{\nu}\bigr)  \star \bigl( D_{\mu}(B) N_{\nu}\bigr)\right) 
\ \ , 
}
together with the corresponding action for the ghost fields, 
\EQ{
S_{\rm ghost} = -2 \int d^{4} x \ \Tr \left(  
\bar{c} \star D_\mu(B)\star D_\mu (B+N) \star c \right) \ \ . 
}
Importantly,
when the background gauge field $B_\mu$ is not held fixed,
the gauge-fixed action $S_{\rm YM} + S_{\rm g.f.} + S_{\rm ghost}$
is gauge-invariant    
under gauge transformations of the background field $B_\mu$   
and rotations of $N_\mu$ (and the ghosts), \ie\ 
\beqa  
\label{btran} 
B_\mu &\longrightarrow &B_\mu^\Omega = \Omega (B_\mu + \partial_\mu ) 
\Omega^{-1} \ \ , 
\\ 
N_\mu &\longrightarrow &N_\mu^\Omega = \Omega N_\mu  \Omega^{-1} \ \ , 
\\ 
c &\longrightarrow &c^\Omega = \Omega\, c\, \Omega^{-1} \ \ , 
\eeqa 
where $\Omega(x)$ is an element of the noncommutative gauge group,  
$\Omega = e^{\alpha^A t^A }_{\star}$, \cite{Ferrara}.

In a one-loop computation of the effective action,  
one discards linear terms in the fluctuations and keeps track of the
quadratic terms.  
After a little algebra we find that the  
gauge-fixed noncommutative Yang-Mills action functional can be rewritten as 
\beqa 
\label{sgf} 
S_{\rm YM} + S_{\rm g.f.} + S_{\rm ghost}& =& - {1 \over 2g^2} 
\int d^{4} x \ \Tr \left( F_{\mu \nu}^B\star  F_{\mu \nu}^B \right)  
\\ \nonumber   
&& -{1 \over g^2} 
\int d^{4} x \ \Tr \left( N_{\mu} \star [ M_{\mu \nu}^{\rm g.f.} , N_\nu ]_{\star}\right) -
2 \int d^{4} x \ \Tr \left( \bar{c} \star D^2(B) \star c \right)   
\ \ , 
\eeqa 
where  
\EQ{
M_{\mu \nu}^{\rm g.f.} = -D^2 (B) \delta_{\mu \nu} - 2 F_{\mu \nu}^B 
\ \ . 
}
In this derivation, we have repeatedly used the fact that  
$ 
\int d^{4} x \ f_1 \star f_2 \star \cdots \star f_{n-1} \star f_n   =  
\int d^{4} x \  f_n\star   f_1 \star \cdots \star f_{n-1} 
$.  
Note that for any field $\phi$ transforming in the adjoint of the gauge group   
(as $N_\mu$ or $c$), \ie\ 
$\phi^\Omega = \Omega \phi \Omega^{-1}$, 
the background covariant derivative 
is given by  $D_\mu (B) \phi = \partial_\mu \phi + [B_\mu , \phi]_{\star}$, and
\EQ{
\label{D2A} 
D^2(B) \star \phi \equiv   
\p^2 \phi + [(\p_\mu B_\mu ) , \phi]_\star + 2[B_\mu , \p_\mu \phi ]_\star  
+  [B_\mu , [B_\mu ,\phi  ]_\star ]_\star 
\ \ . 
}
 
After the functional integration  
over the $N$-fields (and the ghosts) in \eqref{Wdef}  
we are left with an effective action of the $B$ fields only.
Noncommutative gauge-invariance (\ref{btran}) is crucial in  
constraining the interactions which can be generated  
in this procedure. Consequently the 
Wilsonian effective action will always contain the `kinetic term' 
\EQ{
S_{\rm Weff} [ B ] \ni -{1  \over 2g^2_{eff}} 
\int d^{4} x \ \Tr \left( F_{\mu \nu}^B\star  F_{\mu \nu}^B \right)   
\ \ , 
}
where the multiplicative coefficient on the right hand side is identified
with the Wilsonian coupling constant at the corresponding momentum scale.
Of course, in addition to this term there are 
higher-dimensional operators\footnote{It will soon turn out that in
{\it non-supersymmetric} noncommutative theories the
Wilsonian effective action also
contains operators of lower dimension than $F_{\mu \nu}^B\star  F_{\mu \nu}^B$.
These are  non-local operators, 
but perfectly consistent with  noncommutative gauge
invariance. These operators will be singular and unsuppressed in the infrared,
leading to a rather abnormal behaviour of the theory \cite{Matusis}.
}, which are suppressed  
by inverse powers of the cutoff.

We can now generalize our analysis by adding to the pure gauge action
generic matter fields: 
complex scalars  and Weyl fermions  
transforming in general representations of the gauge group\footnote{For 
a discussion of the allowed representations   
of the noncommutative $U(1)$ group see 
\cite{Hayakawa,Ferrara,Terashima:2000xq,Gracia-Bondia}.}, 
which are described by the action
\EQ{
S_{\rm fermi} + S_{\rm scalar} =  
-2 \int d^{4} x \   
\Tr \left( \bar{\lambda}\  \bar{\sigma}_{\mu}\star (D_\mu \lambda)\right) -  
2\int d^{4} x \ \Tr \left( \overline{(D_\mu \phi)}\star(D_\mu \phi)\right)  
\ \ . 
}
 
In what follows we will be interested in the term in the Wilsonian effective
action quadratic in the background gauge field.
For the purpose of determining the 
Wilsonian coupling constant it is sufficient  
to concentrate on the kinetic term $(\p_\mu B_\nu - \p_\nu B_\mu)^2$ 
in the effective Lagrangian, which in momentum space gives  
$2B_\mu (k) B_\nu (-k) (k^2 \delta_{\mu\nu}-k_\mu k_\nu )$.  
In the effective theory the tree level transverse tensor 
$(k^2 \delta_{\mu\nu}-k_\mu k_\nu )$ 
will in general be replaced by a tensor $\Pi_{\mu \nu}(k)$ which from now on
will be referred to as the {\it Wilsonian polarization tensor}.
It is defined so that the term in $S_{\rm Weff}$ quadratic in the background
gauge field is
\EQ{
2 \int {d^4 k \over (2\pi)^4} B_\mu (k) B_\nu (-k)   
\ \Pi_{\mu \nu} 
\ \ . \label{wptdef} 
}

In the effective theory arising from an underlying  
{\it commutative} theory, $\Pi_{\mu \nu}$ has the same tensor structure  
$f(k^2)(k^2 \delta_{\mu\nu}-k_\mu k_\nu )$ as at the classical level
due to Lorentz and gauge invariance.
Only the scalar function
$f(k^2)$ would receive contributions from perturbation theory.  
However, in noncommutative $\Rq$ there is another linearly-independent rank-2 symmetric transverse tensor: 
$\tilde{k}_\mu\tilde{k}_\nu /\tilde{k}^4 $,  
where we have defined $\tilde{k}_\mu =  \theta_{\mu \nu}k_\nu$. It is transverse,  
$k\cdot \tilde{k}\tilde{k}_\nu /\tilde{k}^4
=0=\tilde{k}_\mu\tilde{k}\cdot k /\tilde{k}^4$, since  $\theta_{\mu\nu}$ is antisymmetric.  
On general grounds, the gauge-boson polarization tensor
has the structure
\EQ{
\Pi_{\mu \nu} (k) =  
\Pi_1 (k^2, \tilde{k}^2) (k^2 \delta_{\mu\nu}-k_\mu k_\nu ) +  
\Pi_2 (k^2, \tilde{k}^2) {\tilde{k}_\mu\tilde{k}_\nu \over \tilde{k}^4} 
\ \ . \label{pimunu}
}
The new term $\tilde{k}_\mu\tilde{k}_\nu /\tilde{k}^4 $ has the derivative
dimension $-2$, it is leading compared to the standard gauge-kinetic term
(which has the derivative dimension $+2$), and
singular in the infrared. We will show in Section 5
that $\Pi_2$ vanishes
for all supersymmetric noncommutative gauge theories
(unbroken and softly broken), as was first discussed   
in \cite{Matusis}. 
We will also see that $\Pi_1$ receives contribution from planar 
as well as from  
nonplanar diagrams, whereas $\Pi_2$ is an intrinsically noncommutative object  
and arises only from nonplanar diagrams.  

One-loop computations in the background field method for a standard commutative
theory  literally are a textbook exercise \cite{Peskin}.
To compute $\Pi_{\mu \nu}$ in noncommutative theories we will use the same
elegant bookkeeping approach as in \cite{Peskin}.

Let us introduce the action functional which describes the dynamics of  
a generic spin-$j$ noncommutative field in the representation  
{\bf r} of the gauge group in the background of $B_\mu$: 
\beqa 
S [ \phi ] &=&  
-\int d^{4} x \  \phi_{m,a} \star  
\left ( - D^2 (B)\delta_{mn}\delta^{ab} + 2i (F_{\mu \nu}^B)^{ab}  
\hf J^{\mu \nu}_{mn}   
\right)\star \phi_{n,b} 
\cr 
&\equiv&-\int d^{4} x \   
\phi_{m,a} \star [ \Delta_{j,{\bf r}}]_{mn}^{ab}\star \phi_{n,b} 
\ \ .  
\eeqa 
Here $a,b$ are indices of the  representation {\bf r} of the gauge group,   
$F^{ab}\equiv \sum_{A=1}^{N^2} F^A t^A_{ab}$, $m,n$ are spin indices and  
$J^{\mu \nu}_{mn}$ are  
the generators of the euclidean Lorentz group appropriate  
for the spin of $\phi$: 
\beqa 
J &=& 0 \qquad \qquad \qquad \qquad {\rm for\ spin\ 0 \ fields}, 
\\ \nonumber 
J^{\mu \nu}_{\rho \sigma}&=&i(\delta^\mu_\rho \delta^\nu_\sigma - 
\delta^\nu_\rho\delta^\mu_\sigma) \qquad {\rm for\ 4-vectors}, 
\\ \nonumber 
[J^{\mu \nu}]_{\alpha}^{\ \beta} &=& 
i  \hf [\sigma^{\mu \nu}]_{\alpha}^{\ \beta}
\qquad \qquad {\rm for\ Weyl\ fermions}\ \ . 
\eeqa 
Note that for fields transforming  in the fundamental representation
of the gauge group 
the appropriate covariant derivative is  
$D_\mu (B) \varphi = (\p_\mu + B_\mu)\star \varphi$, so that $D^2 (B) \varphi$ 
reads  
\EQ{
\label{D2F} 
 D^2 (B) \star \varphi =  
\left( \p^2  + (\p_\mu B_\mu) + 2B_\mu \p_\mu + B_\mu \star B_\mu  
\right) 
\star \varphi 
\ \ .   
}
It then follows that, at one-loop level,    
\EQ{
S_{\rm Weff} [B] =  
-{1  \over 2g^2} 
\int d^{4} x \ \Tr F_{\mu \nu}^B\star  F_{\mu \nu}^B - 
\sum_{j,{\bf r}} \alpha_{j} \log {\rm det}_{\star} \Delta_{j, {\bf r}}  
\ \ , 
}
where the sum is extended to all fields in the theory, including ghosts and gauge fields.  
$\alpha_{j}$ is equal to $+1$ ($-1$) for ghost (scalar) fields
and to $+1/2$ ($-1/2$) for 
Weyl fermions (gauge fields). 
For the functional star-determinants we have
\SP{\log {\rm det}_{\star} \Delta_{j, {\bf r}} \equiv&
\log {\rm det}_{\star} (-\partial^2 + {\cal K}(B)_{j, {\bf r}})\\
&= \log {\rm det}_{\star} (-\partial^2) + 
{\rm tr}_{\star} \log (1+(-\partial^2)^{-1}{\cal K}(B)_{j, {\bf r}}) \ .
\label{logkdef}
}
The first term on the second line of \eqref{logkdef} will be dropped
as it is $B$-independent and contributes only to the vacuum loops. 
The second term on the last line of \eqref{logkdef} has an expansion
in terms of Feynman diagrams. 
The next two sections are devoted to the computation  
of these diagrams.
 
\section{Feynman rules}

For simplicity from now on we will restrict our attention
to the noncommutative $U(1)$ gauge group 
(modifications needed for the general $U(N)$ case will be considered elsewhere).
To make our conventions similar to standard QED we change the normalization of the $U(1)$ generator to $t^0 = 1/i$.
We still use the anti-hermitean generator in order to keep
$i$'s out of the rest of the formulae.

\subsection{Adjoint representation} 
Using (\ref{D2A})  we rewrite $\Delta_{j, {\bf r}}$  
acting on adjoint fields $\phi$ as   
\SP{
\Delta_{j, {\bf \sst G}}\star \phi &\equiv 
-\partial^2 \phi + {\cal K}(B)_{j, {\bf \sst G}}\star \phi \\
&=-\partial^2 \phi -  \left[ (\p_\mu B_\mu ) , \phi \right]_\star -  
2 \left[ B_\mu \p_\mu , \phi \right]_{\star}-  
\left[ B_\mu ,  \left[ B_\mu , \phi \right]_\star  \right]_{\star}  
+ 2i \left( \hf J^{\mu \nu}  
\left[ F_{\mu \nu}^B , \phi \right]_{\star} 
\right) 
\ \ .  \label{deladj}
}
The Taylor expansion of the corresponding logarithm in \eqref{logkdef}
will involve the Feynman diagrams made from the three interaction vertices.
The first one is the $\phi$-$B$-$\phi$ vertex 
(which follows from the second and the third terms
on the second line in \eqref{deladj}), 
\beqa  
\label{ncsqued1} 
- \int d^4x \ \bar{\phi} \star  
\left[ (\p_\mu B_\mu ) + 2 B_\mu \p_\mu , \phi \right]_{\star}&=& 
\int {d^4 p' \over (2\pi)^4}   
{d^4 q \over (2\pi)^4}   
{d^4 p \over (2\pi)^4} \  
(2\pi)^4 \delta^{(4)} (p+q-p')\    
\nonumber \\
&&\bar{\phi}(p')  B_\mu (q) \phi (p)  
\left[ - 2 (p+p')_{\mu} \sin {1\over 2} q \tilde{p} \right] 
\ \ . 
\eeqa 
The second vertex $\phi$-$B$-$B$-$\phi$
follows from the fourth term
on the second line in \eqref{deladj},
\beqa 
\label{ncsqued2} 
- \int d^4x \ \bar{\phi} \star  
\left[ B_\mu , \left[ B_\mu , \phi \right]_\star \right]_{\star} 
&=&  
\int {d^4 p' \over (2\pi)^4}   
{d^4 q_1 \over (2\pi)^4}  {d^4 q_2 \over (2\pi)^4}   
{d^4 p \over (2\pi)^4} \  
(2\pi)^4 \delta^{(4)} (p+q_1+ q_2-p')\  
\nonumber \\
&&\bar{\phi}(p')  B_\mu (q_1)  B_\nu (q_2)\phi (p)  
\left[ - 4 \delta_{\mu \nu} 
\sin {1\over 2} p' \tilde{q_1} \ \sin {1\over 2} q_2 \tilde{p} \right] 
\ \ , 
\eeqa 
and the third vertex follows from the last term
on the second line in \eqref{deladj}
\beqa  
\label{ncsqued3} 
i\int d^4x \ \bar{\phi} J^{\mu \nu} \star  
\left[ \p_\mu B_\nu - \p_\nu B_\mu , \phi \right]_{\star} &=& 
\int {d^4 p' \over (2\pi)^4}   
{d^4 q \over (2\pi)^4}   
{d^4 p \over (2\pi)^4} \  
(2\pi)^4 \delta^{(4)} (p+q-p')\   
\nonumber \\
&&\bar{\phi}(p')  J^{\mu \nu} B_\nu (q) \phi (p)  
\left[ 4i q_\mu \sin {1 \over 2} q \tilde{p} \right] 
\ \ . 
\eeqa 
Notice that the interaction term arising from the commutator term  
in $F_{\mu \nu}^B$, 
\beqa 
\label{noncontr1} 
i\int d^4x \ \bar{\phi} J^{\mu \nu} \star  
\left[ \left[ B_\mu , B_\nu \right]_{\star},  \phi \right]_{\star} &=& 
\int {d^4 p' \over (2\pi)^4}   
{d^4 q_1 \over (2\pi)^4}  {d^4 q_2 \over (2\pi)^4}   
{d^4 p \over (2\pi)^4} \  
(2\pi)^4 \delta^{(4)} (p+q_1+ q_2-p')\    
\nonumber \\
&&\bar{\phi}(p')  J^{\mu \nu}  
B_\mu (q_1)  B_\nu (q_2)\phi (p)  
\left[ 4i   
\sin {1\over 2} q_1 \tilde{q_2} \ \sin {1\over 2} p \tilde{p'} \right] 
\ \  
\eeqa 
does not contribute,  to leading order,  
to the vacuum polarization tensor,  
since (\ref{noncontr1}) vanishes when $q_1 + q_2 = 0$. 

The first two vertices (\ref{ncsqued1}) and (\ref{ncsqued2}) 
are the standard Feynman vertices for  
noncommutative electrodynamics with an adjoint scalar field and were
first considered in \cite{Arefeva}. The third expression
\eqref{ncsqued3} is the new, so-called $J$-vertex, specific
to the background field method \cite{Peskin}.

We now move on to fields in the fundamental representation.

\subsection{Fundamental representation} 
We use (\ref{D2F})  and  rewrite $\Delta_{j, {\bf r}}$  
acting on fundamental fields $\varphi$ as  
\EQ{
\Delta_{j, {\bf f}}\star \varphi =  
\left\{  
-\partial^2 -  \left (\p_\mu B_\mu ) + 2 B_\mu \p_\mu + B_\mu \star B_\mu \right)  
+2i \left( {1 \over 2} J^{\mu \nu}  F_{\mu \nu}^B \right)\right\} \star \varphi  
\ \ . 
}
The three vertices are then given by the following expressions:  
\beqa  
\label{fncsqued1} 
- \int d^4x \ \bar{\varphi} \star  
\left[ (\p_\mu B_\mu ) + 2 B_\mu \p_\mu   \right]\star \varphi&=& 
\int {d^4 p' \over (2\pi)^4}   
{d^4 q \over (2\pi)^4}   
{d^4 p \over (2\pi)^4} \  
(2\pi)^4 \delta^{(4)} (p+q-p')\    
 \nonumber \\
&&\bar{\varphi}(p')  B_\mu (q) \varphi (p)  
\left[-  i (p+p')_{\mu} e^{- {i\over 2} q \tilde{p}} \right] 
\ \ , 
\eeqa 
\beqa 
\label{fncsqued2} 
- \int d^4x \ \bar{\varphi} \star  
\left( B_\mu \star B_\mu \right)\star  \varphi  
&=&  
\int {d^4 p' \over (2\pi)^4}   
{d^4 q_1 \over (2\pi)^4}  {d^4 q_2 \over (2\pi)^4}   
{d^4 p \over (2\pi)^4} \  
(2\pi)^4 \delta^{(4)} (p+q_1+ q_2-p')\    
\nonumber \\
&&\bar{\varphi}(p')  B_\mu (q_1)  B_\nu (q_2)\varphi (p)  
\left[  - \delta_{\mu \nu} 
e^{- {i\over 2} (-p' \tilde{q_1} + q_2 \tilde{p})} \right] 
\ \ , 
\eeqa 
\beqa  
\label{fncsqued3} 
i\int d^4x \ \bar{\varphi} J^{\mu \nu} \star  
\left( \p_\mu B_\nu - \p_\nu B_\mu \right)\star \varphi  &=& 
\int {d^4 p' \over (2\pi)^4}   
{d^4 q \over (2\pi)^4}   
{d^4 p \over (2\pi)^4} \  
(2\pi)^4 \delta^{(4)} (p+q-p')\    
 \nonumber \\
&&\bar{\varphi}(p')  J^{\mu \nu} B_\nu (q) \varphi (p)  
\left[ -2 q_\mu e^{-{i \over 2} q \tilde{p}} \right] 
\ \ . 
\eeqa 
Similar considerations to those presented before  apply to the  
the term  
\beqa 
\label{noncontr2} 
i\int d^4x \ \bar{\varphi} J^{\mu \nu} \star  
\left[ B_\mu , B_\nu \right]_{\star} \varphi &=& 
\int {d^4 p' \over (2\pi)^4}   
{d^4 q_1 \over (2\pi)^4}  {d^4 q_2 \over (2\pi)^4}   
{d^4 p \over (2\pi)^4} \  
(2\pi)^4 \delta^{(4)} (p+q_1+ q_2-p')\    
\nonumber \\
&&\bar{\varphi}(p')  J^{\mu \nu}  
B_\mu (q_1)  B_\nu (q_2)\varphi (p)  
\left[ 2 
e^{-{i\over 2} p\tilde{p'} } \ \sin {1\over 2} q_1 \tilde{q_2} \right] 
\ \ , 
\eeqa 
which to leading order gives vanishing contribution  
to the vacuum polarization.  Equations
(\ref{fncsqued1}), (\ref{fncsqued2}) and (\ref{fncsqued3}) are the vertices for 
noncommutative electrodynamics with fundamental scalars,
and the new vertex is specific to the background field treatment.

\section{Planar contributions} 
\label{planarsection}

Expanding the logarithm in \eqref{logkdef} to the second order
in the background fields $B_\mu$ gives the 
Feynman graphs shown in Figures 3, 4 and 5. We depict the 
$J$-vertices by a cross.
The UV-divergent integrals will be regularized by using
 dimensional regularization,
and the UV-divergences will be removed with the supersymmetry-preserving
$\overline{\rm DR}$-scheme \cite{Siegel}.

\subsection{Adjoint representation}
We start with the fields in the  
adjoint representation of noncommutative $U(1)$.

\vfil\eject

\epsfxsize=18cm
\centerline{\epsffile{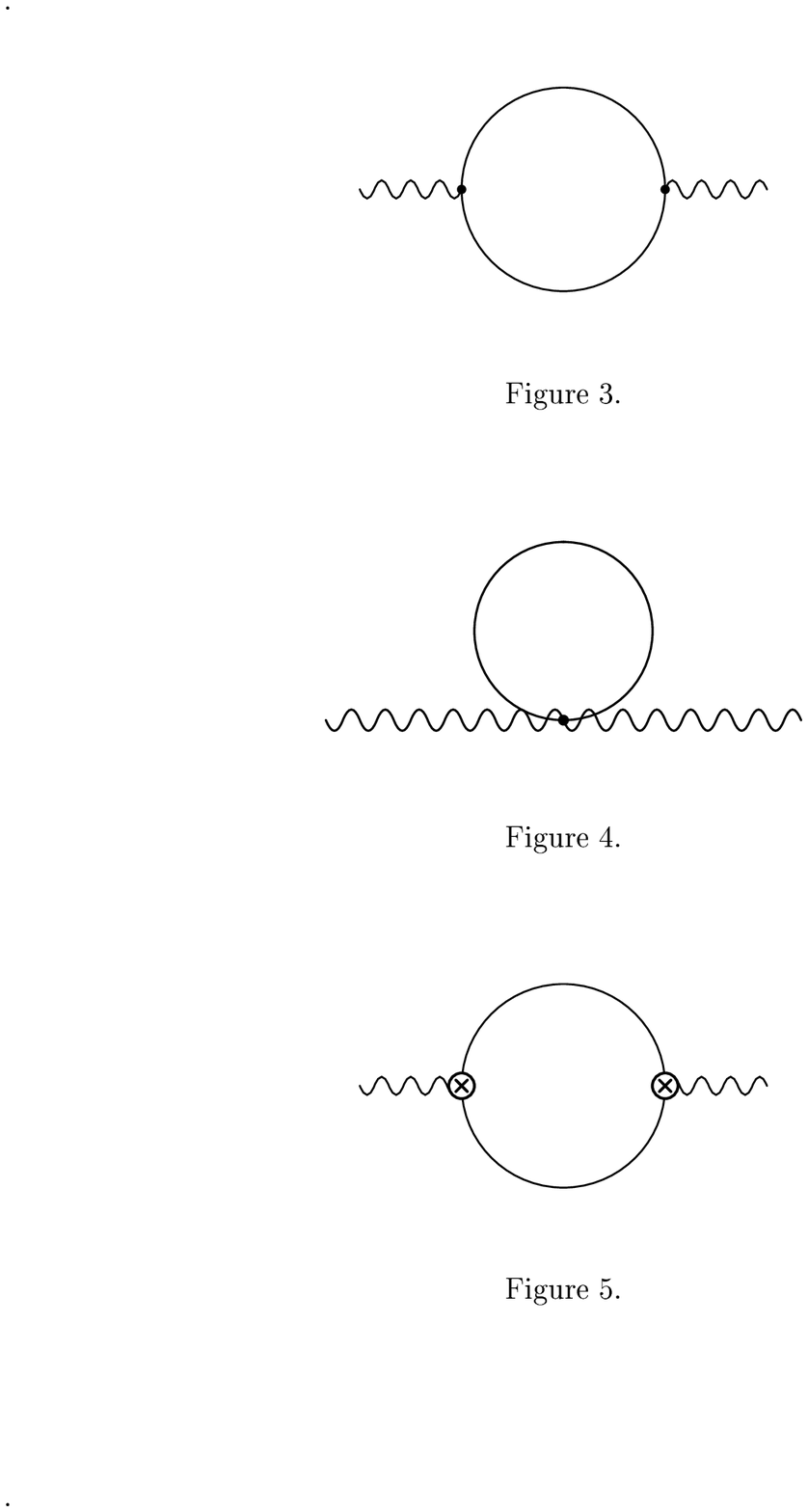}}

\vfil\eject
 
From the first Feynman amplitude (shown in Figure 3) we get  
\EQ{
\label{first} 
-{1 \over 2} \int {d^4 k \over (2\pi)^4} B_\mu (k) B_\nu (-k)    
\int {d^D p \over (2\pi)^D}  
\Tr {-4 \sin^2 {1\over2} k \tilde{p} \ (2p + k)_\mu (2p + k)_\nu \over 
p^2 (p+k)^2} 
\ \ . 
}
The second diagram, Figure 4, gives 
\EQ{
\label{second} 
 \int {d^4 k \over (2\pi)^4} B_\mu (k) B_\nu (-k)    
\int {d^D p \over (2\pi)^D}  
\Tr {- 4 \sin^2 {1\over2}  k \tilde{p}\ \delta_{\mu \nu}\over p^2} 
\ \ , 
}
where the trace is over spin indices. 
Its effect simply amounts to multiply the  
result by 
\EQ{
\Tr \uno_{j} \equiv d(j) 
\ \ , 
}
where $d(j)$ is the number of spin component of the field $\phi$, 
\EQ{
d(j) \equiv \quad 1 \quad
{\rm for\,scalars,}\qquad 2 \quad {\rm for\, Weyl\, fermions,}
\qquad 4 \quad{\rm for\, vectors}. }
 We now examine the  
third amplitude.  It is depicted in Figure 5 and gives  
\EQ{
\label{third} 
-{1\over 2}  \int {d^4 k\over (2\pi)^4} B_\mu (k) B_\nu (-k)    
\int {d^D p \over (2\pi)^D}  
\Tr {-16 J^{\mu \rho} J^{ \nu \lambda}  k_{\lambda} k_{\rho}  
\sin^2 {1\over2}  k \tilde{p} \over p^2 (p+k)^2} 
\ \ , 
} 
where in the spin $j$ representation  
\EQ{
\Tr (J^{\mu \rho} J^{ \nu \lambda})_{j} = C(j) (\delta^{\mu \nu} \delta^{\rho \lambda} -  
\delta^{\mu \lambda} \delta^{\nu \rho}) 
\ \ ,  
}
\EQ{
C(j) \equiv \quad 0 \quad
{\rm for\,scalars,}\qquad \hf \quad {\rm for\, Weyl\, fermions,}
\qquad 2 \quad{\rm for\, vectors}. }

In what follows we will make an extensive use of identity
\EQ{
\label{pnp} 
\sin^2 {1\over2}  k \tilde{p}= {1\over 2} (1-\cos k \tilde{p}) 
\ ,
}
and will refer to the contributions generated by the two 
terms on the right hand side of 
\eqref{pnp} as {\it planar} and {\it non-planar} contributions
respectively.

Planar contributions to $\Pi_{\mu \nu}$
are selected with the substitution 
\EQ{
\sin^2 {1\over2}  k \tilde{p} \longrightarrow {1\over 2} \ ,
}
into the sum of (\ref{first}), (\ref{second}) and (\ref{third}),  
which we write as  
\EQ{
2 \int {d^4 k \over (2\pi)^4} B_\mu (k) B_\nu (-k)   
\ \Pi_{\mu \nu}^{planar} (k; j, {\bf G}) 
\ \ , 
} 
where  
\beqa 
\nonumber 
\Pi_{\mu \nu}^{planar} (k; j, {\bf G})&=&  
{1\over 2}\int {d^4 p \over (2\pi)^4} \left\{ \left[   
 { (2p + k)_\mu (2p + k)_\nu  
\over 
p^2 (p+k)^2} -  
{ 2 \delta_{\mu \nu}  \over p^2}\right] C({\bf G}) d(j) 
+4 {k^2\delta_{\mu \nu} - k_{\mu} k_{\nu} \over  
p^2 (p+k)^2}C({\bf G}) C(j)\right\} 
\\ \nonumber \\  
&=& \left( [\Pi_{\mu \nu}^{planar}]_{1,2} +   
 [\Pi_{\mu \nu}^{planar}]_{3}\right) 
(k; j, {\bf G})  
\ \ .  
\eeqa 
Here  $[\Pi_{\mu \nu}^{planar}]_{1,2}$  denotes the planar contributions  
arising from the sum of the first two Feynman diagrams, and  
$[\Pi_{\mu \nu}^{planar}]_{3}$ arises from the third one. 
Performing the $D$-dimensional integral, we get  
\EQ{
\label{div12} 
[\Pi_{\mu \nu}^{planar}]_{1,2} (k; j, {\bf G}) =  
{1\over 2}(k^2 \delta_{\mu \nu} - k_\mu k_\nu)\left\{ - {d(j) \over (4\pi)^2}  
\left[ {1 \over 3} \left( {2\over \epsilon} - \gamma_{E} \right)  
- \int_{0}^{1} dx \ (1-2x)^2 \log {A(k^2 , x) \over 4\pi \mu^2} \right]\right\}  
\ \ , 
}
where $A(k^2 ,x)= k^2 x(1-x)$, and we used   
$\Gamma (2 - {D \over 2}) = \Gamma ({\epsilon \over 2}) =  
{2\over \epsilon} - \gamma_E + O(\epsilon )$, where  $\gamma_E$ is  the  
Euler-Mascheroni constant. The mass parameter $\mu$ is introduced to keep  
the coupling constant dimensionless in $4-\epsilon$ dimensions, redefining 
$g\rightarrow  g \mu^{\epsilon / 2}$. 
 
We now compute the  planar part of the third Feynman graph, which  
leads to   
\EQ{
\label{div3} 
[\Pi_{\mu \nu}^{planar}]_{3} (k; j, {\bf G}) =  
{1\over 2} 
(k^2 \delta_{\mu \nu} - k_\mu k_\nu ) \left\{ {4 C(j) \over (4\pi)^2}  
\left[ {2\over \epsilon} - \gamma_{E}   
- \int_{0}^{1} dx \ \log {A(k^2 , x) \over 4\pi \mu^2} \right]\right\}  
\ \ . 
} 
From (\ref{div12}) and (\ref{div3}) we extract the divergent part of the  
planar contribution,   
\EQ{
\label{sum} 
\Pi_{\mu \nu}^{div} (k; j, {\bf G})=  
- {1\over 2} (k^2 \delta_{\mu \nu} - k_\mu k_\nu )  
\left[ {{d(j)\over 3}  - 4 C(j) \over (4\pi)^2} 
\right]  
\cdot \left( {2\over \epsilon} \right)  
\ \ . 
}

\subsection{Fundamental representation}
We now discuss the case of fundamental matter.  
It turns out that in this case 
the exponential factors   
mutually cancel in  the first and third diagram,  
whereas  become 1 in  the second. In other words 
only the planar contribution is present%
\footnote{This was firstly seen in noncommutative electrodynamics  
in \cite{Hayakawa2}.}.     
More precisely, by comparing (\ref{ncsqued1}), (\ref{ncsqued2}), (\ref{ncsqued3})  
to (\ref{fncsqued1}), (\ref{fncsqued2}), (\ref{fncsqued3})  
respectively, one immediately   realizes that  
the expressions for the Feynman graphs for fundamental matter are simply obtained 
from (\ref{first}), (\ref{second}), (\ref{third})  
by making the substitution  
\EQ{
4 \sin^2 (p \tilde{q} / 2) \longrightarrow 1 
\ \ . \label{fpldef}
}
Recalling (\ref{pnp}), we conclude that the planar contribution  
from the adjoint field $\phi_{j}$  is exactly {\it twice} the one  for matter fields 
in the fundamental, $\varphi_j$, whereas nonplanar contributions vanish.  
 
\subsection{The microscopic {\bf $\beta$} function} 
As usual, from (\ref{sum}) one can read
the running of the renormalized coupling constant as a function
of the subtraction point $\mu$, which in turn gives the $\beta$ function
of the microscopic theory,
$\beta (g) = [\mu {\p \over \p \mu} g]_{g_0 , \epsilon}$.
This way we get    
\beqa 
\label{beta1} 
\beta (g) &=& - {b_0 \over 16\pi^2}g^3 + O(g^5) \ , \\  
b_0 &=& 2 \sum_{j; {\bf r}={\bf f},{\bf G}} 
\alpha_j  \left[ {1\over 3} d(j) - 4 C(j) \right] C ({\bf r}) 
\ , \label{betatwo} 
\eeqa 
where we have formally {\it defined}  
\beqa  
\label{cvalues} 
C ({\bf G}) = 1 \ \ , &  
C ({\bf f}) = {1 \over 2} 
\ \ . 
\eeqa 

We stress that this $\beta$ function
is defined in a mass-independent regularization scheme and does not change
when massive particles are decoupled. The running of the Wilsonian coupling
constant is correctly reproduced by it only in the extreme UV region.

The expression for $b_0$ in (\ref{beta1}) can be recast in a more familiar  form  
by noticing that the quantity $(1/3) d(j) - 4 C(j)$ is equal to $1/3$, $-4/3$, $-20/3$ 
for complex scalars, Weyl fermions and gauge fields respectively. 
If we call $n_{\bf r_f}$ ($n_{\bf r_b}$) the number of Weyl fermions  
(complex scalars)  transforming in  
the representation ${\bf r_f}$ (${\bf r_b}$), we get, summing over all  
the fields in the theory,   
\EQ{
\label{mitte} 
b_0 =  
2 \left( {11 \over 3}C ({\bf G}) - 
{2 \over 3}\sum_{\bf f} n_{\bf r_f}C ({\bf r_f}) - 
{1 \over 3}\sum_{\bf b}n_{\bf r_b}C ({\bf r_b}) 
\right)     
\ \ . 
}
The overall factor of 2 which multiplies (\ref{mitte}) is merely  
a consequence of the choice of the $U(1)$ generator.  
Had we normalized it to $1/2$, as usually done for the $SU(N)$ generators  
in ordinary commutative Yang-Mills theories, 
this factor of 2 would not have been present.  
 
\subsection{Supersymmetric theories}
Two observations are in order.   
First, notice that in supersymmetric theories there is a precise cancellation
between bosonic and fermionic degrees of freedom,
\EQ{
\label{susyzero} 
\sum_{j} \alpha_j d(j) = 0 \ \ ,\ \  {\rm for\  any\ {\bf r}\ of} \ G 
\ \ . 
} 
Second, let us explore in which theories
the microscopic $\beta$ function vanishes.  
Using  (\ref{susyzero}) and \eqref{betatwo} we see that in this case  
\EQ{
\sum_{j,{\bf r}} \alpha_{j} C({\bf r}) C(j) =0
\ ,  \label{n4zero}
}
which means 
\EQ{
C ({\bf G}) \left( 4 - {\mathcal N}\right) = C ({\bf f})\  N_{\bf f} 
\ \ , 
}
where ${\mathcal N}$ is the number of supersymmetries and $N_{\bf f}$ 
the number of  fundamental hypermultiplets.
Therefore, 
pure ${\mathcal N}=4 $ theories and ${\mathcal N}=2$ theories with 
$N_{\rm f}=2$  fundamental hypermultiplets 
have vanishing microscopic $\beta$ function.
 
Before closing this section, let us write down the full planar contribution  
to $\Pi_{\mu \nu}$ for a generic supersymmetric theory.  
First we define the scalar function $\Pi^{planar} (k^2)$ via  
\beqa 
\label{planarsusy} 
\Pi_{\mu \nu}^{planar} (k) &=& (k^2 \delta_{\mu \nu} - k_\mu k_\nu ) \Pi^{planar} (k^2) 
\\ \nonumber  
&\equiv & \sum_{j, {\bf r}} \alpha_{j}\Pi_{\mu \nu}^{planar} (j, {\bf r})   
\ \ , 
\eeqa 
where the sum is  over all fields in the theory.   
Then, using (\ref{susyzero}) and the definitions 
given in (\ref{cvalues}) we finally get    
\EQ{
\label{planarsusy2} 
\Pi^{planar} (k^2) =  
{2 \over (4\pi )^2 }\left( \sum_{j, {\bf r}} \alpha_{j} C(j) C ({\bf r}) \right) 
 \left[ {2\over \epsilon} - \gamma_{E}   
- \int_{0}^{1} dx \ \log {A(k^2 , x) \over 4\pi \mu^2} \right]  + O(\epsilon)
\ \ . 
}
We will make use of this expression in Section 6.

\section{Nonplanar contributions and IR/UV mixing} 
\label{nonplanarsection} 
 
As we have seen from the discussion preceeding \eqref{fpldef},
nonplanar contributions arise only from fields in the adjoint representation
of noncommutative $U(1)$. They are read off from
(\ref{first}), (\ref{second}) and (\ref{third}) with the substitution
\EQ{
\sin^2 {1\over2}  k \tilde{p} \longrightarrow -{1\over 2} e^{{i\over2}  
k \tilde{p}}\ .
}
These contributions
are finite, so that we can work directly in $D=$4 dimensions and write
\EQ{
\label{firstnp} 
2 \int {d^4 k \over (2\pi)^4} B_\mu (k) B_\nu (-k)   
\ \Pi_{\mu \nu}^{np} (k; j, {\bf G}) 
\ \ , 
} 
where  
\beqa
\Pi_{\mu \nu}^{np} (k; j, {\bf G})=  
&{1\over 2}& \int {d^4 p \over (2\pi)^4} \left( \left[   
 {- (2p + k)_\mu (2p + k)_\nu  
\over 
p^2 (p+k)^2}  +
{ 2 \delta_{\mu \nu}  \over p^2}\right] C({\bf G}) d(j) \right.
\nonumber \\ 
&-& 4   \left.
{k^2\delta_{\mu \nu} - k_{\mu} k_{\nu} \over  
p^2 (p+k)^2}C({\bf G}) C(j)\right) e^{i p\tilde{k}} 
\ \ . 
\eeqa
Like in (\ref{pimunu}), let us decompose $\Pi_{\mu \nu}^{np} (k )$  as   
\EQ{
\label{pimununp} 
\Pi_{\mu \nu}^{np} (k ) =  
\Pi_1^{np} (k^2, \tilde{k}^2) (k^2 \delta_{\mu\nu}-k_\mu k_\nu ) +  
\Pi_2^{np} (k^2, \tilde{k}^2) {\tilde{k}_\mu\tilde{k}_\nu \over \tilde{k}^4} 
\ \ . 
}
It is also convenient to introduce the quantities 
\beqa  
\hat{\Pi} = \delta_{\mu \nu}[\Pi^{np} (k)]^{\mu \nu}   & \ \ , &  
\tilde{\Pi} =  {\tilde{k}_{\mu}  \tilde{k}_{\nu}\over \tilde{k}^2} 
[\Pi^{np} (k) ]^{\mu \nu} 
\ \ , 
\eeqa 
which are related to $\Pi_1^{np}$, $\Pi_2^{np}$ via 
\beqa 
\Pi_1^{np} = { 1\over 2 |k|^2}\left(\hat{\Pi} - \tilde{\Pi}\right) & \ \ , & 
\Pi_2^{np}=  {\tilde{k}^2 \over 2 }\left(-\hat{\Pi} + 3\tilde{\Pi}\right) 
\ \ . 
\eeqa 
Using 
\beqa 
\int {d^4 p \over (2\pi)^4}  
{e^{i p\tilde{k}}\over p^2 (p+k)^2}&=&{2 \over (4\pi)^2} \int_{0}^{1}dx \  
K_{0} (\sqrt{A} |\tilde{k}|) 
\ \ , 
\\ 
\int {d^4 p \over (2\pi)^4}  
{e^{i p\tilde{k}}\over p^2 }&=& 
{1 \over (4\pi)^2}{4\over \tilde{k}^2} 
\ \ , 
\eeqa 
where $A= k^2x(1-x) $, one finds 
\beqa 
\hat{\Pi} &=& {C({\bf G}) \over (4\pi)^2}\left\{ 
{8 d(j) \over \tilde{k}^2} - k^2\left[ 12C(j) - d(j)\right] 
\int_{0}^{1}dx \  
K_{0} (\sqrt{A} |\tilde{k}|)\right\} 
\ \ , 
\\ 
\tilde{\Pi}& =& {4C({\bf G})\over (4\pi)^2}\left\{ 
{ d(j) \over \tilde{k}^2}-  \left(C(j) k^2 -  
d(j) {\tilde{k}_\mu \tilde{k}_\nu\over \tilde{k}^2} 
{\p \over \p \tilde{k}_\mu }  {\p \over \p \tilde{k}_\nu }\right)  
 \int_{0}^{1}dx \  
K_0 (\sqrt{A} |\tilde{k}|)\right\} 
\ \ . 
\eeqa 
The Bessel function $K_0(z)$ has an expansion 
\EQ{
K_0 (z) = -\log {z\over 2}  \left[ 1 + {z^2 \over 4} + O(z^4)\right]  
-\gamma_{E} - (\gamma_{E}-1){z^2 \over 4} + O(z^4) 
\ \ ,  
}
which leads to the following expressions for  
$\Pi_1^{np}$, $\Pi_2^{np}$:  
\beqa 
\label{loga} 
&\Pi_1^{np}& (k^2, \tilde{k}^2; j, {\bf G}) \simeq 
\nonumber \\
&&- {C({\bf G}) \over (4\pi)^2}\left\{ 
\left( {d(j)\over 3}  - 4 C(j) \right)  
\left(\log {|k| |\tilde{k}| \over 2} + \gamma_E \right) 
+ {1\over 2} \left[ 8 C(j) - {5\over 6} d(j)\right]\right\} 
\ \ , 
\\ 
&\Pi_2^{np}& (k^2, \tilde{k}^2; j, {\bf G}) 
\simeq - {C({\bf G}) \over (4\pi)^2}\left\{-8d(j) + {d(j)\over 4}  
k^2 \tilde{k}^2 
\right\} 
\ \ . 
\eeqa 
Notice that the $1/\tilde{k}^2$ pole has exactly cancelled.  
We are now ready to  
sum over all fields in the theory, and find the complete  
nonplanar contribution to the vacuum polarization,  
\EQ{
\Pi_{\mu \nu}^{np} = \sum_{j} \alpha_{j}\Pi_{\mu \nu}^{np} (j, {\bf G})  
\ \ . 
}

Recall  from (\ref{pimununp}) that  $\Pi_2$ is  
the coefficient in front of $\tilde{k}_\mu\tilde{k}_\nu /\tilde{k}^4 $. 
This term has the  same dimension  
as $(\p_\mu B_\nu - \p_\nu B_\mu )^2$, but it   
diverges quadratically as $\tilde{k}\rightarrow 0$. It gauge invariant,  
and it is not surprising that it has been generated.

In the case of supersymmetric theories, there is an important simplification,
which follows from
(\ref{susyzero}): the contributions arising from  
$\Pi_2$ {\it sum up to zero}, as first observed in  
\cite{Matusis}.   
Therefore, summing over all fields
we get compact exact expressions 
\beqa 
\label{nonplanarsusy} 
\Pi_1^{np}(k^2, \tilde{k}^2)&=&  - {4 C({\bf G}) \sum_{j} \alpha_j C(j) \over (4\pi)^2} 
\int_{0}^{1}dx \  
K_0 (\sqrt{A} |\tilde{k}|) 
\ \ , 
\\ \label{nppp}
\Pi_2^{np}(k^2, \tilde{k}^2)&=&  0 
\ \ . 
\eeqa 

Notice that a logarithmic singularity is generated 
in (\ref{loga}) whenever  
$k$ (or $\theta$) go to zero.  
However, this singularity is not present for ${\mathcal N}=4$
since in this case the sum over $j$ in the expression above is zero. 
These theories  have vanishing $\beta$ function, and  
it is then natural to ask  
whether the logarithmic singularity 
also vanishes in ${\mathcal N}=2$ theories  
with two fundamental hypermultiplets, which have vanishing microscopic
$\beta$ function as well.  
The answer to this question is no. In fact, no nonplanar contribution is generated  
by fundamental hypermultiplets circulating in loops, so the contribution  
coming from the ${\mathcal N}=2$ adjoint superfield is not cancelled.

Finally, note that 
for  $ {{\mathcal N}=4}$ theories \eqref{n4zero} implies that
both the planar and the nonplanar contribution 
to the vacuum polarization vanish:  
\beqa 
\left[ \Pi_{\mu \nu}^{planar}\right]_{{\mathcal N}=4} = 0 \ \ , &  
\left[ \Pi_{\mu \nu}^{np}\right]_{{\mathcal N}=4} = 0 \ \ . 
\eeqa

 
\section{Flowing down from ${\mathcal N=4}$ and finite ${\mathcal N=2}$} 
\label{applications} 

As an application of the Wilsonian approach to noncommutative gauge theories 
outlined in the previous sections,  
we now study how the decoupling of heavy degrees of freedom occurs in a  
noncommutative setup.  
In the previous analysis, we have found two examples of theories that  
are finite and free from quadratic infrared divergences: 
${\mathcal N}=4$ pure supersymmetric Yang-Mills and   
${\mathcal N}=2$ supersymmetric Yang-Mills  with two
fundamental hypermultiplets. 
In the former, we can give mass to two of the four (${\mathcal N}=1$) chiral  
superfields, thus breaking $\N=4$ supersymmetry to $\N=2^*$. 
In the latter we give mass
to the fundamental hypermultiplets. 
Naively
the effective theories obtained  
in the two cases by integrating out   
massive degrees of freedom  would be expected to belong to  
the same universality class in the extreme IR. To explore this issue
we examine  
the behaviour of the Wilsonian coupling constants  
as functions of the momentum scale as this scale is lowered down
from the UV-region to the infrared.  

We start with the $\N=2^*$ theory. Our previous construction outlined
in Sections \ref{planarsection} and \ref{nonplanarsection}
is modified by the presence of non-zero mass $m$ for the two adjoint
hypermultiplets. In general,   
the contribution to  $\Pi_{\mu \nu}$ coming from a {\it massive} 
particle circulating in loops  
is taken into account by simply making in  
(\ref{planarsusy2}) and  (\ref{nonplanarsusy})  
the substitution  
\EQ{
A (k^2, x) \longrightarrow 
A(k^2, x; m^2_{j, {\bf r}}) \equiv k^2 x (1-x) +  m^2_{j, {\bf r}} 
\ \ , 
}
where $m_{j, {\bf r}}$ is the mass of a spin $j$ particle belonging  
to the representation {\bf r} of the gauge group. 
 This in turn implies that for ${\mathcal N}=2^*$ theories $\Pi_{\mu \nu}$  
no longer vanishes; instead one has  
\beqa 
\label{41} 
\Pi_1^{np} &=& - {4C({\bf G})  \over (4\pi)^2} \left( \sum_{j} C(j) \alpha_j \right)  
  \int_{0}^{1}dx \left[   
K_{0} ( \sqrt{k^2 x(1-x) + m^2} |\tilde{k}|) -  
K_{0} ( \sqrt{k^2 x(1-x)} |\tilde{k}|)\right] 
\nonumber
\\ 
\label{42} 
\Pi_2^{np} &=& 0 
\ \ , \nonumber
\\ 
\label{43} 
\Pi_{planar}& =& - { 2 C({\bf G})  \over (4\pi)^2}   
\left( \sum_{j} C(j) \alpha_j \right)  
\int_{0}^{1}dx \ \log {m^2 + k^2 x(1-x)  \over k^2 x(1-x)} 
\ \ , 
\eeqa 
where the sum is over massive particles only.  
Therefore  $\sum_{j} C(j) \alpha_j  C({\bf G}) = (1/2)\cdot (1/2)\cdot 2=1/2$. 
Notice that, as anticipated in the Introduction, no IR-singularities appear 
in the nonplanar part, as in the massless case.

The Wilsonian coupling at the scale $k^2$ is  given by  
\EQ{
\label{44} 
{1\over 4 g_{\N=2^*}^2(k^2)} = {1\over 4 g_{\rm micro}^2} + \Pi_{\N=2^*} (k^2 ) 
\ \ , 
}
so that we conclude that  
\beqa 
\label{45} 
{1\over  g_{\N=2^*}^2(k^2)} &=& {1\over  g^2_{\rm micro}} 
+ {[b_0]_{{\mathcal N}=2^\star} \over (4\pi)^2}  
\left\{ \int_{0}^{1}dx \ \log {k^2 x(1-x)  \over m^2 + k^2 x(1-x)} + \right. 
\nonumber \\ 
&+& \left. 
2  \int_{0}^{1}dx \left[   
K_{0} ( \sqrt{k^2 x(1-x) } |\tilde{k}|) -  
K_{0} ( \sqrt{k^2 x(1-x)+ m^2 } |\tilde{k}|)\right] \right\} 
\ \ , 
\eeqa 
where we have used (\ref{mitte}) to compute  
$[b_0]_{{\mathcal N}=2^\star}= 2\cdot (11/3 - 4/3 -1/3) = 4$. 
Note that the right hand side of \eqref{45} goes to 
$1/g_{\rm micro}^2$ as $k^2\rightarrow 0$.
The running of the Wilsonian coupling
can be conveniently described via
\EQ{{1\over  g_{\N=2^*}^2(k)} = {1\over  g^2_{\rm micro}} + {1 \over 2 \pi^2}
f_{\N=2^*}(k,\theta^{-1/2},m) \ . \label{fn2ast}
}
The function $f_{\N=2^*}(k,\theta^{-1/2}=10^1,m=10^3)$ is shown in Figure 6.

\vfil\eject

\epsfxsize=18cm 
\centerline{\epsffile{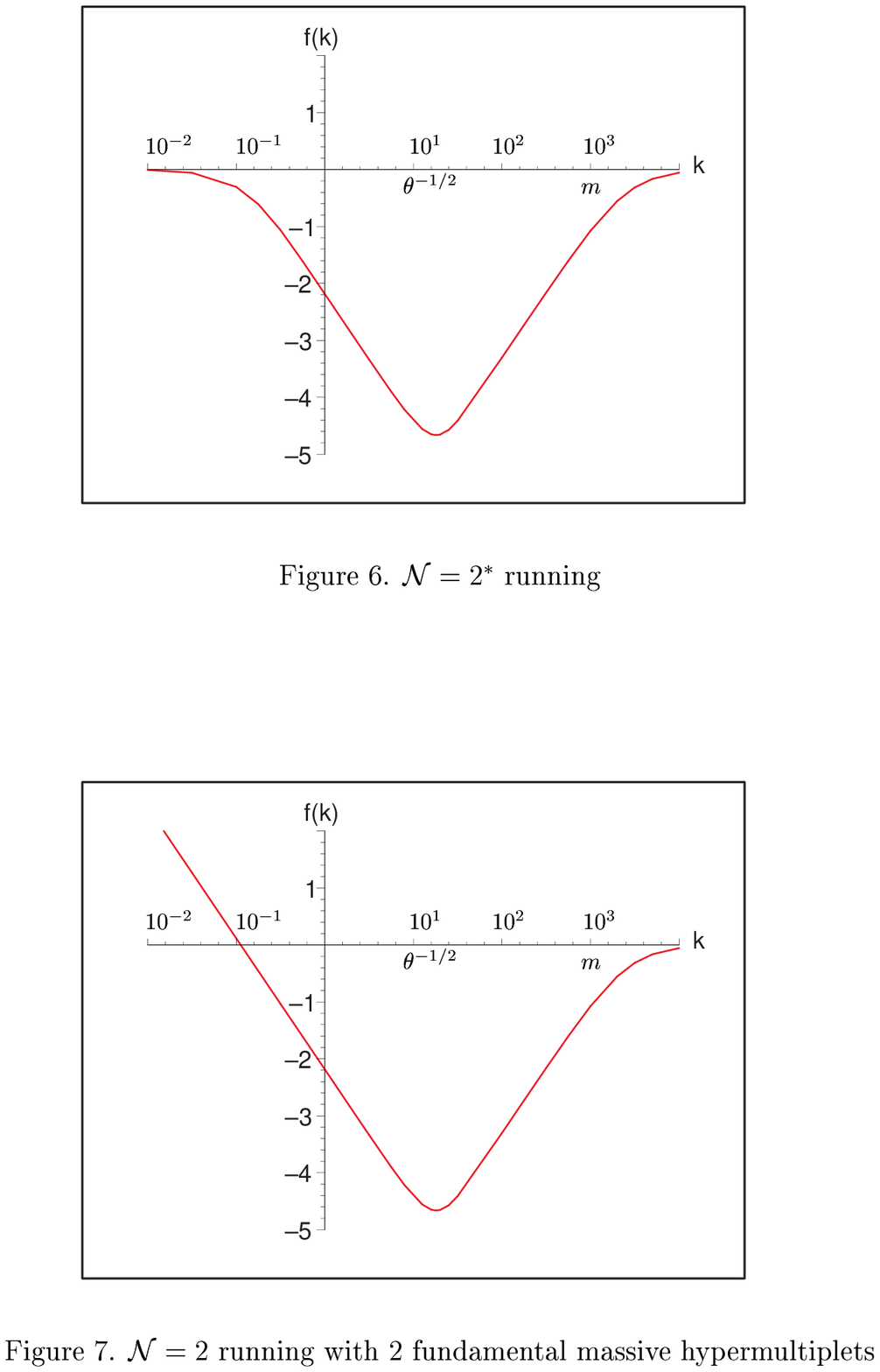}}

\vfil\eject

Now we repeat the same analysis starting from ${\mathcal N}=2$ 
$U(1)$ supersymmetric Yang-Mills 
with two massive hypermultiplets and obtain
\beqa  
\Pi_1^{np} &=& {2  \over (4\pi)^2}   
\int_{0}^{1}dx \  
K_{0} ( \sqrt{k^2 x(1-x) } |\tilde{k}|)   
\ \ , \nonumber
\\ 
\Pi_2^{np} &=& 0 
\ \ , \nonumber
\\ 
\label{23} 
\Pi_{planar}& =& { 1\over (4\pi)^2}   
\int_{0}^{1}dx \ \log {k^2 x(1-x)  \over m^2 + k^2 x(1-x)} 
\ \ . 
\eeqa 
The Wilsonian coupling at the scale $k^2$ is now given by  
\beqa 
\label{25} 
{1\over  g_{\N=2}^2(k^2)} &=& {1\over  g_{\rm micro}^2} 
+ {[b_0]_{{\mathcal N}=2} \over (4\pi)^2}  
\left[ \int_{0}^{1}dx \ \log {k^2 x(1-x)  \over m^2 + k^2 x(1-x)} + \right. 
\nonumber \\ 
&+& \left. 
2  \int_{0}^{1}dx \  
K_{0} ( \sqrt{k^2 x(1-x) } |\tilde{k}|) \right]  
\ \ , 
\eeqa 
where again, in the normalizations of (\ref{mitte}),    
$[b_0]_{{\mathcal N}=2}= 4$. 
Finally, we represent the running of the Wilsonian coupling via
\EQ{{1\over  g_{\N=2}^2(k)} = {1\over  g^2_{\rm micro}} + {1 \over 2 \pi^2}
f_{\N=2}(k,\theta^{-1/2},m) \ , \label{fn2}}
and plot the function $f_{\N=2}(k,\theta^{-1/2}=10,m=10^3)$ in Figure 7.
As anticipated earlier $f$ exhibits a logarithmic behaviour in the infrared.

\section{Wilsonian flow in asymptotically free 
supersymmetric theories} 

Our general formalism can be instantly applied to asymptotically free 
supersymmetric gauge theories. In this Section we briefly analyze
the evolution of the Wilsonian coupling constant 
in pure ${\cal N}=2$ and ${\cal N}=1$  Super Yang-Mills theories. 

The Wilsonian coupling at the scale $k^2$ is  given by  
\EQ{
\label{2121} 
{1\over 4 g_{\rm eff}^2(k^2)} = {1\over 4 g_{\rm micro}^2} + \Pi (k^2 ) 
\ \ , 
}
where $\Pi=\Pi^{planar} +\Pi_1^{np}$. The renormalized  
planar contribution $\Pi^{planar}$ is obtained from \eqref{planarsusy2} by
subtracting the pole in $\epsilon$ in the $\overline{\rm DR}$ scheme,
\EQ{
\label{planagain} 
{1\over 4 g_{\rm micro}^2} +
\Pi^{planar} (k^2) = - 
{2 \over (4\pi )^2 }\left( \sum_{j, {\bf r}} \alpha_{j} C(j) C ({\bf r}) \right) \int_{0}^{1} dx \ \log {k^2 x(1-x) \over \Lambda^2_{\sst \overline{\rm DR}}} 
\ \ . 
}
The nonplanar contribution was calculated in \eqref{nonplanarsusy}. 
Putting these results together, we obtain the running of the Wilsonian
coupling,  
\EQ{
\label{4512} 
{1\over  g_{\rm eff}^2(k^2)} = 
{b_0 \over (4\pi)^2} \left(\log{k^2\over \Lambda_{\sst \overline{\rm DR}}^2}
-2\right) 
+ {2b_0 \over (4\pi)^2}
\int_{0}^{1}dx\,   
K_{0} ( \sqrt{ x(1-x) } |k||\tilde{k}|)
\ \ , 
}
where $b_0$ is the first coefficient of the $\beta$ function given by
(\ref{mitte}). In our $U(1)$ normalization  
$[b_0]_{{\mathcal N}=2}= 4$ and $[b_0]_{{\mathcal N}=1}= 6.$
From this we can easily read the UV and the IR limits
of the $1/g_{\rm eff}^2$:
\beqa
1\over  g_{\rm eff}^2 &\rightarrow&
{b_0 \over (4\pi)^2} \log k^2 \ , \qquad {\rm as} \ 
k^2\to\infty \ ,
\\ 
{1\over  g_{\rm eff}^2} &\rightarrow&
-{b_0 \over (4\pi)^2} \log k^2 \ , \qquad {\rm as} \
k^2\to 0 \ .
\eeqa 
This shows that for not too large values of the microscopic coupling 
$g_{\rm micro}$ the effective coupling 
$g_{\rm eff}$ does not diverge ($1\over g_{\rm eff}^2$ does not reach zero).
We saw that this was the case for the finite theories analyzed in the previous
section. Remarkably this still holds in asymptotically free theories.

\medskip

We conclude with the observation that standard holomorphy arguments
\cite{Seiberg} imply that the Wilsonian coupling in $\N=2$ theories
is in fact one-loop exact in perturbation theory.

\section*{Acknowledgements} 
We would like to thank Nigel Glover, Gian Carlo Rossi and Massimo Testa for  
discussions. 
G.T. was supported by a PPARC SPG grant 
and the Angelo Della Riccia Foundation.

\vfil\eject

\startappendix

\Appendix{The NSVZ $\beta$ function for $U(N)$} 
The $\beta$ function was computed for pure Yang-Mills theory with $U(1)$ group 
in \cite{Sheikh-Jabbari,Martin},  for pure $U(N)$ in  
\cite{Minwalla,Krajewski,Armoni} and for QED in \cite{Hayakawa} with ordinary  
perturbation theory. In Section 4.3
we have extended these results adding  
adjoint as well as fundamental matter for the gauge group $U(1)$.  
We now present an argument which leads 
to a generalization of the previous result  
(\ref{beta1}) for supersymmetric  $U(N)$ gauge theories.  
We will show that $b_0$ is still given  
by (\ref{mitte})%
\footnote{With the factor of 2 deleted,  
we return to the normalization $\Tr (t^A t^B ) = - \delta^{AB} / 2$,  
$A,B = 1,\ldots ,N^2$}  
with the only replacement  
\EQ{ 
C ({\bf G}) = N 
\ \ . 
}
The derivation is a simple extension of the well-known  
Novikov-Shifman-Vainshtein-Zakharov (NSVZ) derivation of the $\beta$ function   
in \cite{Novikov:1983,Novikov:1984,Novikov:1986}, so we will only sketch the argument.  
We limit our attention to pure supersymmetric gauge theories, and implicitly work  
with holomorphic gauge couplings (for recent discussions on the subtleties related to  
the definition of canonical and holomorphic gauge couplings  
and the presence of matter fields,  
we refer the reader to  \cite{Dine,Arkani-Hamed:1998,Arkani-Hamed:2000}) 
The first observation is that noncommutative gauge theories have instanton configurations,  
even in the case of gauge group $U(1)$ 
\cite{Nekrasov:1998ss}. In the NSVZ derivation of the exact $\beta$ function, 
a key element was the renormalizability of the theory to all orders in perturbation theory.  
At present there is still no proof that supersymmetric theories are renormalizable,  
but strong evidence is supported by one-loop perturbative calculations; in \cite{Arefeva} 
the one-loop renormalizability of ${\mathcal N}=2$ supersymmetric Yang-Mills  
theories on noncommutative on $\Rq$ was proven,  
while in \cite{Girotti,Bichl} it was shown that  
the Wess-Zumino model is renormalizable to all orders.  
We will then {\it assume} their renormalizability.  
In the calculation of instanton-dominated  
Green's functions, determinants of the kinetic operators  
of the quantum fluctuations around the instanton configuration cancel due to the  
d'Adda-Di Vecchia theorem \cite{D'Adda},  
which is also  at work in the noncommutative case.    
Then the result of a typical one-instanton amplitude is simply  
proportional to
\EQ{
\label{nomu} 
M^{n_B - {1\over 2}n_F}\  e^{-{8\pi^2 \over g^2}} 
\ \ . 
}
Here $M$ is the Pauli-Villars mass scale which is usually introduced in supersymmetric  
instanton calculus, and  $n_B$ ($n_F$) is the number of bosonic (fermionic)  
zero-modes of the Dirac operator 
on noncommutative $\Rq$ in the background of a charge one $U(N)$ instanton.  
In \cite{Nekrasov:1998ss} it was shown that  
the moduli space of noncommutative $U(N)$ instantons is given by a resolution of singularities of 
the ordinary moduli space of commutative instantons  (for  
recent reviews, see \cite{Nakajima,brief}).  
In the case at hand, we have precisely $n_B=4N$. Supersymmetry  relates 
this number to $n_F$, giving  $n_F ={1\over 2}  {\mathcal N} n_B$ where ${\mathcal N}$  is  
the number of supersymmetries; we conclude that  
\EQ{
n_B - {1\over 2}n_F = 4N\left( 1-{{\mathcal N}\over 4}\right) 
\ \ . 
}
Under the stated assumptions, $M$ and $g=g(M)$ in (\ref{nomu}) conspire to give a  
renormalization  group invariant expression,  
\EQ{
M {d \over dM} \left( M^{n_B - {1\over 2}n_F}\  e^{-{8\pi^2 \over g^2 (M)}} \right)= 0 
\ \ , 
}
from which it immediately follows that  
\EQ{
\label{vanishes} 
b_0 \left[ U(N)\right] =   
4N\left(1-{{\mathcal N}\over 4}\right) 
\ \ . 
}
Notice that (\ref{vanishes}) vanishes for  ${\mathcal N}=4$, as in the commutative case.

\end{document}